\documentclass[10pt]{IEEEtran}
\usepackage{amsmath,amsthm,amssymb}
\usepackage{graphicx,epstopdf,epsfig}
\usepackage{float}
\usepackage[caption=false,font=footnotesize]{subfig}
\usepackage{enumitem}
\usepackage{xcolor}

\newtheorem{theorem}{Theorem}

\newtheorem{proposition}[theorem]{Proposition}
\newtheorem{assumption}[theorem]{Assumption}

\allowdisplaybreaks

\def\x{\boldsymbol{x}}
\def\y{\boldsymbol{y}}

\def\q{\boldsymbol{q}}
\def\A{\mathbf{A}}
\def\W{\mathbf{W}}
\def\D{\mathbf{D}}
\def\K{\mathbf{K}}
\def\P{\mathbf{P}}
\def\Q{\mathbf{Q}}
\def\I{\mathbf{I}}
\def\prox{\mathrm{prox}}

\def\argmin{\mathop{\mathrm{argmin}}}

\begin{document}

\title{Plug-and-play ISTA converges with kernel denoisers}
\author{Ruturaj~G.~Gavaskar,~\IEEEmembership{Student~Member,~IEEE}, and Kunal~N.~Chaudhury,~\IEEEmembership{Senior~Member,~IEEE}
\thanks{Dept. of Electrical Engineering, Indian Institute of Science, Bengaluru, India. K.~N.~Chaudhury was supported by SERB Grant SB/S3/EECE/281/2016 from DST, Government of India. Correspondence: ruturajg@iisc.ac.in.}
}

\maketitle

\begin{abstract}
Plug-and-play (PnP) method is a recent paradigm for image regularization, where the proximal operator (associated with some given regularizer) in an iterative algorithm is replaced with a powerful denoiser. Algorithmically, this involves repeated inversion (of the forward model) and denoising until convergence. Remarkably, PnP regularization produces promising results for several restoration applications. However, a fundamental question in this regard is the theoretical convergence of the PnP iterations, since the algorithm is not strictly derived from an optimization framework. This question has been investigated in recent works, but there are still many unresolved problems. For example, it is not known if convergence can be guaranteed if we use generic kernel denoisers (e.g. nonlocal means) within the ISTA framework (PnP-ISTA). We prove that, under reasonable assumptions, fixed-point convergence of PnP-ISTA is indeed guaranteed for linear inverse problems such as deblurring, inpainting and superresolution (the assumptions are verifiable for inpainting).
We compare our theoretical findings with existing results, validate them numerically, and  explain their practical relevance.
\end{abstract}


\section{Introduction}
\label{sec:intro}

Restoration problems such as denoising, deblurring, inpainting, and superresolution \cite{Gunturk2012_img_restoration} are often modeled as optimization problems of the form
\begin{equation}
\label{eq:main_prob}
\min_{\x \in \mathbb{R}^n} \ \ f(\x) + \lambda g(\x),
\end{equation}
where $\x$ is the image variable, $f(\x)$ is the data term involving the forward model, and $g(\x)$ is the regularizer (possibly induced by some ground-truth prior). For Gaussian denoising, $f(\x) = (1/2) \lVert \x - \y \rVert_2^2$, where $\y = \x_0 + \boldsymbol{n}$ is the noisy image, $\x_0$ is the clean image, and $\boldsymbol{n}$ is white Gaussian noise.
The minimizer of \eqref{eq:main_prob} in this case (i.e. the denoised image)  is technically referred to as the proximal map of $\lambda g$ at $\y$ \cite{Rockafellar2006_prox_map}:
\begin{equation}
\label{eq:prox}
\prox_{\lambda g} (\y) = \argmin_{\x \in \mathbb{R}^n} \ \ \frac{1}{2} \lVert \x - \y \rVert_2^2 + \lambda g(\x).
\end{equation}
Different choices of $g$ result in different denoising schemes, such as total-variation and wavelet denoising \cite{Afonso2010_SALSA}. 
However, not all denoising methods are originally conceived as proximal maps. For example, powerful denoisers such as NLM \cite{Buades2005_NLM} and BM3D \cite{Dabov2007_BM3D} are based on patch-based filtering, while DnCNN \cite{Zhang2017_DnCNN} uses deep neural networks. These denoisers are typically more powerful than total-variation and wavelet denoisers that are derived from a regularization framework. A natural question thus is can we exploit the excellent denoising capability of NLM and BM3D for regularization purpose? One such approach leads to the so-called Plug-and-Play (PnP) method \cite{Venkatakrishnan2013_PnP,Sreehari2016_PnP}. This is based on the observation that 
most standard algorithms \cite{Lions1979_DRS,daubechies2004iterative,Chambolle2011_PD,Boyd2011_ADMM} for solving \eqref{eq:main_prob} such as ISTA and ADMM involve the iterative inversion of the forward model followed by 
the application of \eqref{eq:prox}. Based on the formal equivalence of  the proximal map with Gaussian denoising, 
the authors in \cite{Venkatakrishnan2013_PnP,Sreehari2016_PnP} choose to formally replace \eqref{eq:prox} with NLM or BM3D denoiser.
The resulting algorithm essentially involves repeated inversion and denoising. 
Empirically, PnP methods have been shown to yield excellent results for tomography \cite{Sreehari2016_PnP,Sun2019_PnP_SGD}, superresolution \cite{Chan2017_PnP_conv}, single-photon imaging \cite{Chan2017_PnP_conv}, fusion \cite{Teodoro2019_PnP_fusion}, inpainting \cite{Tirer2019_iter_denoising}, and several other applications \cite{Ono2017_PD_PnP,Zhang2017_IRCNN,Meinhardt2017_learning_prox_op,Buzzard2018_PnP_CE,Chan2019_PnP_graph_SP,Teodoro2019_targeted_PnP,Kamilov2017_PnP_ISTA,Nair2019_hyperspectral_fusion}.

One straightforward way of establishing theoretical convergence of a PnP method is to express the associated denoiser as the proximal map of some regularizer. Indeed, this the case with symmetrized NLM and Gaussian mixture denoisers \cite{Sreehari2016_PnP,Teodoro2019_PnP_fusion}, which can be expressed as the proximal map of a convex regularizer $g(\x)$.
The PnP iterations amount to minimizing \eqref{eq:main_prob} in this case and convergence follows from standard optimization theory as a result. However, establishing convergence for NLM, BM3D or neural networks is difficult, since it is not known if these denoisers can be expressed as proximal maps. Several recent works have addressed the question of theoretical convergence of PnP methods
\cite{Sreehari2016_PnP,Chan2017_PnP_conv,Sun2019_PnP_SGD,Dong2018_DNN_prior,Teodoro2019_PnP_fusion,Tirer2019_iter_denoising,Ryu2019_PnP_trained_conv}.
Fixed-point convergence has been established in \cite{Chan2017_PnP_conv,Ryu2019_PnP_trained_conv}, but the underlying assumptions are often strong or not verifiable. For example, the denoiser is assumed to be bounded in a certain sense  \cite{Chan2017_PnP_conv} which is difficult to check even for NLM, and $f$ is assumed to be strongly convex  \cite{Ryu2019_PnP_trained_conv} which excludes inpainting and superresolution. We also note that convergence  properties of the PnP iterates have been investigated in \cite{Sun2019_PnP_SGD,Tirer2019_iter_denoising,Dong2018_DNN_prior}.

In a different direction, theoretical convergence was established  in \cite{Sreehari2016_PnP}, although using a modified form of NLM called DSG-NLM (whose cost is about three times of NLM  \cite{Unni2018_PnP_LADMM}). More precisely, DSG-NLM  can be expressed as a linear map $\x \mapsto \W \x$, where $\W$ is doubly stochastic, symmetric, and positive semidefinite. Using these properties, it was shown that $\W$ can be expressed as a proximal map, which settled the convergence question for DSG-NLM. However, this and the other papers do not address the question if convergence can be an issue with original NLM. The present work was motivated by the above and the following related questions:

(i) Can fixed-point convergence be established when $\W$ is standard NLM or a generic kernel denoiser \cite{Milanfar2013_filtering_tour} whose cost is about a third of that than DSG-NLM? That is, can the symmetry and doubly-stochastic assumptions be removed?\\
\indent (ii) Can the strong convexity assumption \cite{Ryu2019_PnP_trained_conv} on $f(\x)$ be removed (which excludes  inpainting and superresolution)?\\
\indent (iii) Can convergence be established under assumptions that are practically verifiable?

\textbf{Contributions}.
We address the above questions when ISTA \cite{daubechies2004iterative,figueiredo2003algorithm} is the base algorithm, i.e., when the proximal map in ISTA is replaced by a denoiser.
The resulting PnP method (PnP-ISTA) has been investigated in \cite{Meinhardt2017_learning_prox_op,Sun2019_PnP_SGD,Ryu2019_PnP_trained_conv}. In this paper, we establish the following results which are novel to the best of our knowledge: \\
\indent (i) We prove that under reasonable assumptions, PnP-ISTA exhibits fixed-point convergence for linear inverse problems using standard NLM as the denoiser (Theorem \ref{thm:main}). 

\indent (ii) We prove that these assumptions are met for inpainting (Proposition \ref{prop:inpainting_Q}), and we explicitly compute a bound on the step size in PnP-ISTA that can guarantee convergence (Theorem \ref{thm:inpainting_bound}).

\textbf{Organization}. The problem statement is described in \S\ref{sec:problem}.
We state and explain our convergence results, along with the required assumptions, in \S\ref{sec:results} and defer the technical proofs to \S\ref{sec:proofs}.
The results are further discussed and experimentally validated using inpainting experiments in \S\ref{sec:disc}.
Finally, we summarize our findings in \S\ref{sec:conc}.

\section{Problem statement}
\label{sec:problem}

Our objective is to investigate the convergence of PnP-ISTA with generic kernel denoisers, and NLM in particular.
An advantage of ISTA over other iterative algorithms such as ADMM is that the ISTA update rule \cite{daubechies2004iterative,figueiredo2003algorithm} is simple and involves just one variable.
Applied to \eqref{eq:main_prob}, the rule is given by
\begin{equation}
\label{eq:ISTA}
\x_{k+1} = \prox_{\gamma g} \big( \x_k - \gamma \nabla \! f(\x_k) \big), \qquad (k \geqslant 0)
\end{equation}
where $\x_0$ is the initial guess and $\gamma > 0$ is the step size.
We now describe the general settings of the problem that we study in terms of the denoiser used and the data term $f(\x)$.

We assume that the denoiser used in PnP-ISTA is a kernel denoiser \cite{Milanfar2013_filtering_tour} such as NLM \cite{Buades2005_NLM}.
Such a denoiser can be expressed as linear map $\x \mapsto \W \x$, where $\W \in \mathbb{R}^{n \times n}$ is row-stochastic (i.e. each row of $\W$ sums to $1$), but not symmetric.
Recall that NLM denoises an image by performing weighted averaging of pixels in a square window $\Omega_i$ around pixel $i$ \cite{Buades2005_NLM}. We will refer to $\Omega_i$ as the denoising window around $i$.
The weight matrix $\W$ is derived from a symmetric kernel (affinity) matrix $\K$ as $\W = \D^{-1} \K$, where $\D$ is the diagonal matrix of the row-sums of $\K$.
In NLM, $\K$ is computed from image patches and a large window $\Omega_i$ is allowed \cite{Buades2005_NLM}. 

In general, $\K$ may be computed from an image which is different from the input image.
A common practice in PnP methods is to use the image from the previous iteration to compute $\K$ for the current iteration. 
This is done in the first few iterations, and then $\K$ is frozen for the rest of the iterations \cite{Sreehari2016_PnP,Unni2018_PnP_LADMM} .
Hence, after a finite number of iterations, the denoiser acts as a linear map $\x \mapsto \W \x$.
We adopt this practice in PnP-ISTA.

Moreover, we focus on linear inverse problems, where $f(\x) = (1/2) \lVert \A\x - \y\rVert_2^2$ and $\y$ is the observed image.
This class subsumes a wide variety of problems; in particular inpainting, deblurring, and superresolution \cite{Chan2017_PnP_conv,Tirer2019_iter_denoising}.
Substituting $f(\x) = (1/2) \lVert \A\x - \y\rVert_2^2$ in \eqref{eq:ISTA} and replacing $\prox_{\gamma g}(\cdot)$ by $\W(\cdot)$, we can express the PnP-ISTA update as:
\begin{equation}
\label{eq:affine}
\x_{k+1} = \P \x_k + \q,
\end{equation}
where $\P = \W (\I_n - \gamma \A^\top \! \A)$, $\q = \gamma \W \A^\top \! \y$ and $\I_n$ is the $n \times n$ identity matrix.
In other words, the evolution of the iterates is governed by a linear dynamical system.
Note that $\P$ and $\q$ depend on $\gamma$.

Our aim is to provide sufficient conditions under which the sequence $(\x_k)$ generated via \eqref{eq:affine} converges, irrespective of the initialization $\x_0$. This is referred to as fixed-point convergence to distinguish it from other forms of convergence \cite{Chan2017_PnP_conv} such as objective convergence \cite{Sreehari2016_PnP,Teodoro2019_PnP_fusion}. In particular, we wish to guarantee that the PnP iterates do not diverge or oscillate. Note that if $(\x_k) \longrightarrow \x^\star$, then it follows from (4) that $ \x^\star = (\I - \P)^{-1}\q$ (from the proof in Section VI, it follows that $\I - \P$ is invertible since $1$ is not an eigenvalue of $\P$). Since $\P$ and $\q$ depend on $\gamma$, the limit point will change with the step size  $\gamma$.
We next state and discuss our results.

\section{Convergence Results}
\label{sec:results}

A possible line of attack is to use the contraction mapping principle \cite{granas2013fixed}. We write \eqref{eq:affine} as $\x_{k+1} = T(\x_k)$ where $T(\x)=\P \x + \q$, and note that
$\lVert T(\y_1)  - T(\y_2) \rVert_2 \leq \lVert \P \rVert  \lVert \y_1  - \y_2 \rVert_2$ for $\y_1, \y_2 \in \mathbb{R}^n$,
where $\lVert \cdot \rVert_2$ is the Euclidean norm on $\mathbb{R}^n$ and $\lVert \P \rVert$ is the spectral norm of $\P$ (largest singular value). Convergence of $(\x_k)$ would follow if we can show that $\lVert \P \rVert < 1$, since $T$ would be a contraction map in this case. However, note that $\lVert \P \rVert \leqslant \lVert \W \rVert \lVert \I_n - \gamma \A^\top \! \A \rVert.$
Hence, to ensure $\lVert \P \rVert < 1$ we must have $\lVert \I_n - \gamma \A^\top \! \A \rVert < \lVert \W \rVert^{-1}$.
However, in applications such as  inpainting where $\A$ has a non-trivial null space, $0$ is an eigenvalue of $\A^\top \! \A$, whereby $1$ is an eigenvalue of $\I_n - \gamma \A^\top \! \A$.
As a result, $\lVert \I_n - \gamma \A^\top \! \A \rVert \geqslant 1$.
To ensure $\lVert \P \rVert < 1$, we must therefore make $\lVert \W \rVert < 1$.
However, since $\W$ is row-stochastic, one of its eigenvalues is $1$ \cite{Meyer2000_MA}, so that $\lVert \W \rVert \geqslant 1$. As a result, we cannot use the contraction argument.

Fortunately, there exists a weaker condition on $\P$ which implies convergence of $(\x_k)$.
In particular, motivated by the resemblance of \eqref{eq:affine} to a linear dynamical system, we use a standard condition that guarantees the stability of such a system. This is in terms of the spectral radius of $\P$:
\begin{equation*}
\varrho(\P) = \max \Big\{ |\mu| : \mu \text{ is an eigenvalue of } \P \Big\}.
\end{equation*}
\begin{theorem}
\label{thm:conv}
Let $\varrho(\P) < 1$. Then $(\x_k)$ generated by \eqref{eq:affine} converges from any arbitrary initialization $\x_0$.
\end{theorem}
A proof of this standard result can be found in \cite[Chapter 7]{Meyer2000_MA}.
In other words,  if we can ensure that $\varrho(\P) < 1$, then PnP-ISTA is guaranteed to converge.
Note that since $\W$ and $\A$ are fixed, the step size $\gamma$ is the only parameter we can use to manipulate $\varrho(\P)$.
Therefore, the hope is that $\varrho(\P)$ can be made less than $1$  by controlling $\gamma$.
While bounding the spectral radius of $\P$ seems more difficult than bounding its norm, we will prove that this is possible provided we make a set of assumptions as given below.
We define  $\Q = \W \A^\top\! \A$, so that we can write $\P = \W - \gamma \Q$.
\begin{assumption}
\label{asm:main}
The following conditions hold:
\begin{enumerate}[label=\upshape(\roman*),ref=\thetheorem (\roman*)]
\item\label{asm:W} For $1 \leqslant i,j \leqslant n$, $W_{ij} \geqslant 0$ and $W_{ij}=0$ iff $j \notin \Omega_i$.
\item\label{asm:A} For $1 \leqslant i,j \leqslant n$, $A_{ij} \geqslant 0$.
\item\label{asm:Q} For $1 \leqslant i \leqslant n$,
\begin{equation}
\label{eq:Q}
\sum_{j \notin \Omega_i} Q_{ij} < \sum_{j \in \Omega_i} Q_{ij}.
\end{equation}
\end{enumerate}
\end{assumption}
We have used $W_{ij}$ to denote the $(i,j)$th element of $\W$.
Assumption \ref{asm:W} holds for NLM, where the weights are strictly positive within the denoising window.
Assumption \ref{asm:A} holds in the case of inpainting (where the rows of $\A \in \mathbb{R}^{m \times n}$ are a subset of the rows of $\I_n$ with $m < n$ \cite{Tirer2019_iter_denoising}), deblurring (where $\A$ is the circulant matrix corresponding to a nonnegative point spread function \cite{Tirer2019_iter_denoising}), and superresolution (where $\A$ is the product of nonnegative subsampling and blurring matrices \cite{Chan2017_PnP_conv}).
Assumption \ref{asm:Q}  states that the sum of $Q_{ij}$ within the denoising neighborhood $\Omega_i$ is strictly larger that those outside $\Omega_i$.
Although this assumption is a little difficult to interpret, it can in principle be tested based on $\Q$ and $\Omega_i$.
For the special case of inpainting, we can in fact prove that the condition always holds under a mild assumption (see Section \ref{sec:proofs} for the proof).
\begin{proposition}
\label{prop:inpainting_Q}
Suppose Assumption \ref{asm:W} holds and $\A$ is the subsampling (decimation) matrix.
Moreover, suppose that for every pixel $i$, the denoising window $\Omega_i$ contains at least one observed pixel.
Then Assumption \ref{asm:Q} holds.
\end{proposition}
The requirement in Proposition \ref{prop:inpainting_Q} is not very strict, at least for uniform sampling. 
Indeed, we can simply increase $\Omega_i$ until we hit an observed pixel (recall that $\Omega_i$ is usually large in NLM).
In other words, Assumption \ref{asm:Q} is verifiable for the inpainting problem, in the sense that it can be proved to hold.
We are now ready to state our main convergence result. 
\begin{theorem}
\label{thm:main}
Suppose Assumption \ref{asm:main} holds. Then there exists $\delta > 0$ such that PnP-ISTA converges for any  initialization $\x_0$ provided $0 < \gamma < \delta$.
\end{theorem}
\begin{figure}[t!]
\centering
\subfloat[Ground-truth.]{\includegraphics[width=0.30\linewidth]{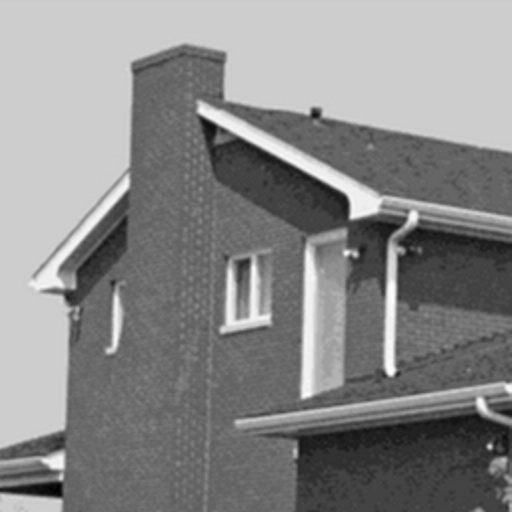}}
\hspace{0.1mm}
\subfloat[Observed pixels.]{\includegraphics[width=0.30\linewidth]{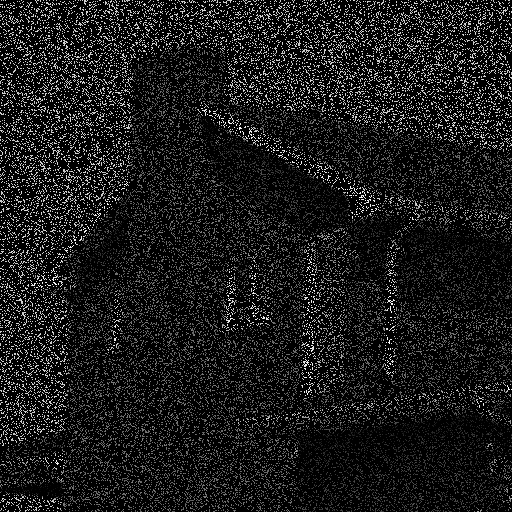}}
\hspace{0.1mm}
\subfloat[Output ($33$ dB).]{\includegraphics[width=0.30\linewidth]{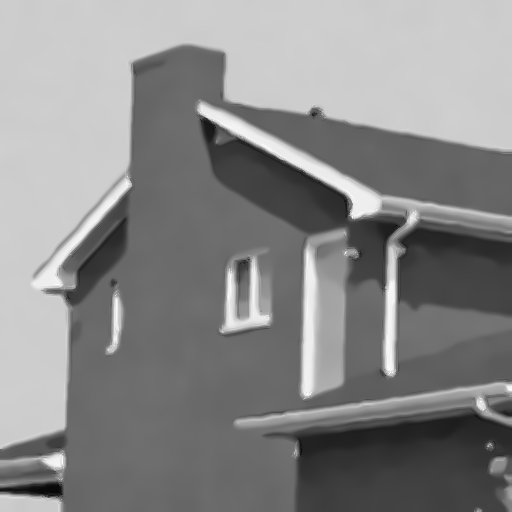}}
\caption{Image inpainting with $80 \%$ missing pixels and  Gaussian noise with standard deviation $10/255$. We used PnP-ISTA with NLM denoiser, where the step size is $\gamma = 0.9$.}
\label{fig:inpainting}
\end{figure}
The proof is given in Section \ref{sec:proofs}, and is based on Theorem \ref{thm:conv}.
An immediate implication of Theorem \ref{thm:main} combined with Proposition \ref{prop:inpainting_Q} is that for inpainting, PnP-ISTA converges if the step size $\gamma$ is sufficiently small.
However, Theorem \ref{thm:main} only affirms the existence of an upper bound; it does not provide a way to compute the bound.
Hence in practice, we can choose a sufficiently small $\gamma$.
However, as stated below, the bound can be explicitly computed for inpainting (see Section \ref{sec:proofs} for the proof). We note that in \cite[Proposition 2]{Sun2019_PnP_SGD}, it is shown that the mean of the squares of the first $k$ residuals in PnP-ISTA decays at $O(1/k)$ rate.
This is different from Theorem \ref{thm:main} in which the convergence of $(\x_k)$ is established.
\begin{theorem}
\label{thm:inpainting_bound}
For the inpainting problem, PnP-ISTA \eqref{eq:affine} converges for any arbitrary initialization if $0 < \gamma < 1$.
\end{theorem}

\section{Discussion}
\label{sec:disc}

\begin{figure}[t!]
\centering
\subfloat[]{\includegraphics[width=0.37\linewidth]{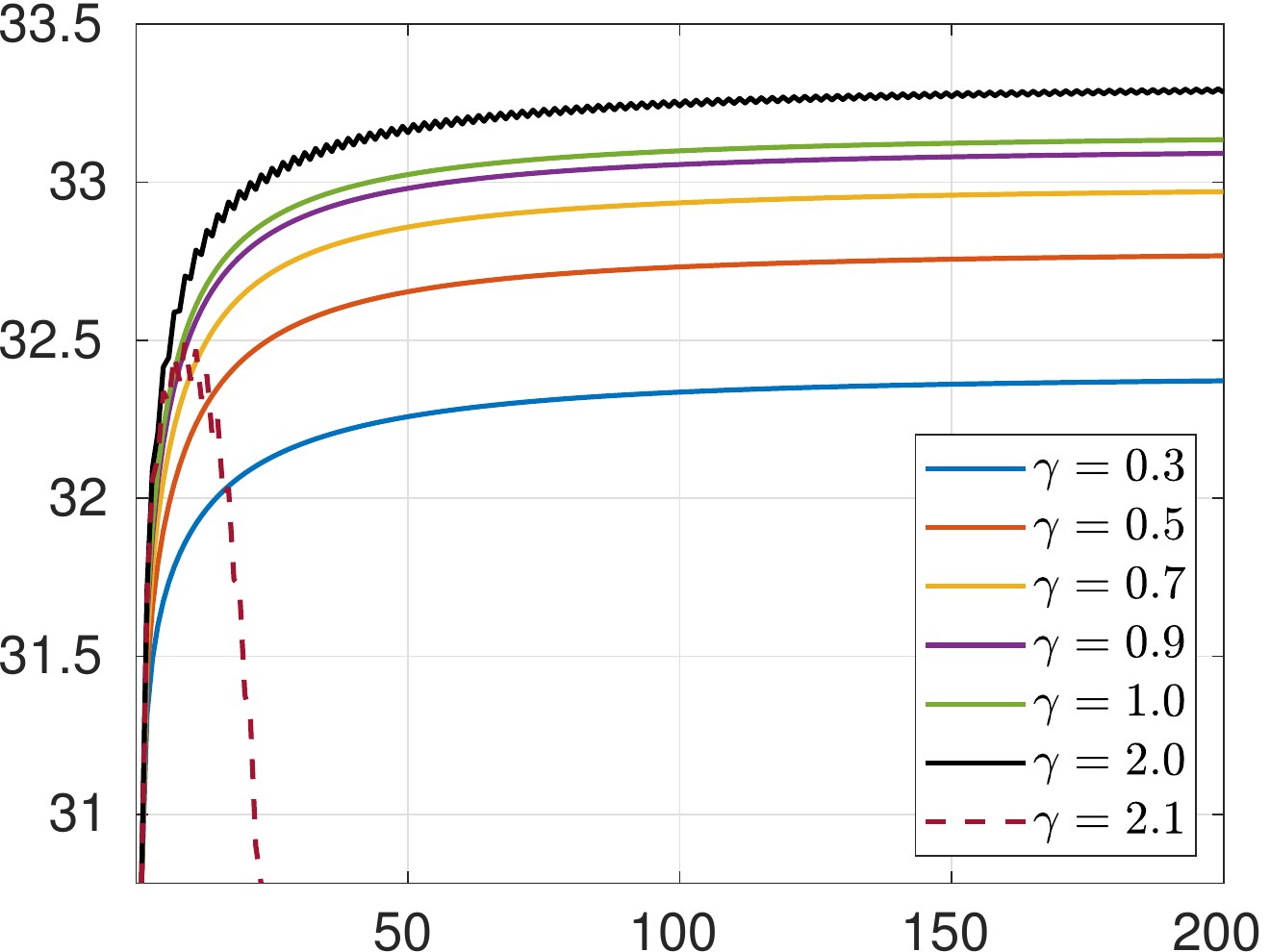}}
\hspace{0.1mm}
\subfloat[]{\includegraphics[width=0.352\linewidth]{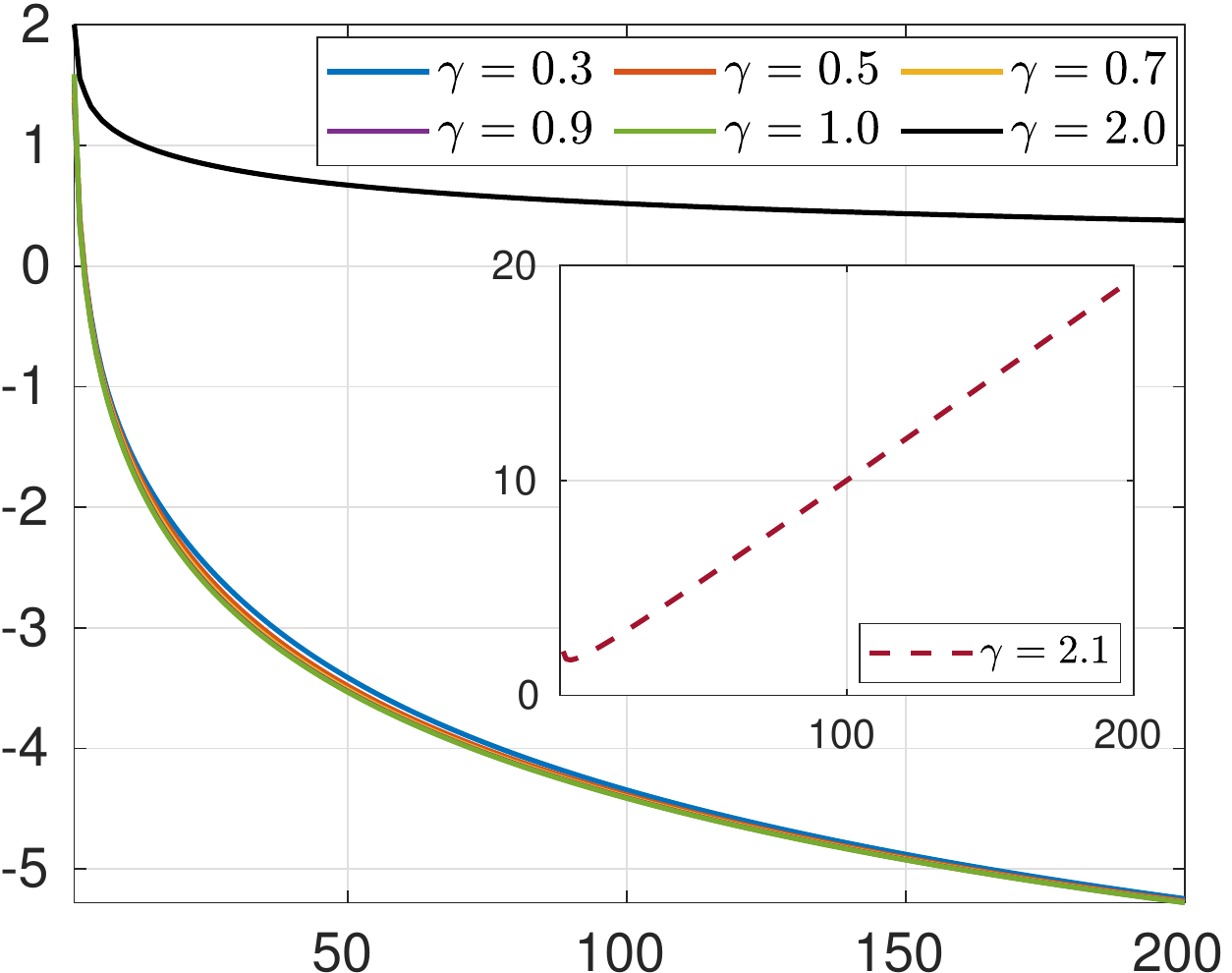}}
\caption{Evolution of (a) PSNR (in dB) and (b) residual (on a log scale) with iterations for different $\gamma$. The settings are as in Figure \ref{fig:inpainting}.}
\label{fig:plots}
\end{figure}

Theorem \ref{thm:inpainting_bound} rigorously establishes the fixed-point convergence of PnP-ISTA for inpainting using the NLM denoiser if we choose a step size $\gamma \in (0,1)$.
To validate this finding, we perform an inpainting experiment in Figure \ref{fig:inpainting} in which just $20 \%$ pixels have been observed. The initialization $\x_0$ is computed by applying a median filter on the input image \cite{Tirer2019_iter_denoising}.
We have used NLM in which the weights are computed from $\x_0$, and we set $\gamma = 0.9$.
The reconstruction in Figure \ref{fig:inpainting}(c) is found to be reasonably accurate.
As expected, for $\gamma \in (0,1)$, we observe that the PSNR between the ground-truth and the iterate $\x_k$ stabilizes as $k$ increases (cf. Figure \ref{fig:plots}(a)).
Moreover, the residuals $\lVert \x_{k+1} - \x_k \rVert_2/255$ go to $0$ as the algorithm progresses (Figure \ref{fig:plots}(b)). This  corroborates with the theoretical result in Theorem \ref{thm:inpainting_bound}.
In fact, the algorithm is empirically found to stabilize even for $\gamma=1$ and $\gamma=2$.
(Note that Theorem \ref{thm:inpainting_bound} does not tell us what happens if $\gamma \geqslant 1$.)
But it fails to stabilize for $\gamma=2.1$.
The corresponding residuals also tends to diverge (see inset in Figure \ref{fig:plots}(b)).

An interesting exercise is to connect the result in Theorem \ref{thm:inpainting_bound} with the convergence of standard ISTA for the inpainting problem.
For convex problems, convergence of ISTA is guaranteed under a couple of assumptions \cite{Beck2009_FISTA}: (i) $\nabla \! f$ is Lipschitz continuous with a Lipschitz constant $L$, and (ii) $0 < \gamma < 1/L$.
For inpainting, the former holds with $L=1$, since the largest eigenvalue of $\A^\top \! \A$ is $1$.
Thus, the objective in \eqref{eq:main_prob} converges to its minimum if $\gamma \in (0,1)$; but proving fixed-point convergence of the iterates is technically challenging.
Note that Theorem \ref{thm:inpainting_bound} proves convergence of the PnP-ISTA iterates using kernel denoisers under the same condition on $\gamma$.
On the other hand, the upper bound on the step size in ISTA can be relaxed from $1/L$ to $2/L$ while still ensuring convergence \cite{Selesnick2010_TV_filtering} (although this no longer ensures a monotonic reduction of the objective). Based on the observation in Figure \ref{fig:plots} that PnP-ISTA stabilizes even for $\gamma = 2$, we conjecture that a similar relaxation of the bound might be possible for PnP-ISTA.

Finally, we note that although Assumption \ref{asm:Q} is difficult to verify for inverse problems other than inpainting, it is empirically observed that PnP-ISTA converges for deblurring with $\gamma \leq 2$.
We refer the reader to the supplement for a couple of related experiments.

\section{Conclusion}
\label{sec:conc}
We proved that PnP-ISTA converges for a class of linear problems using kernel denoisers. The central observation is that the problem of convergence  can be reduced to that of the stability of a dynamical system. Moreover, the assumptions used in this regard can be verified for inpainting and we can obtain a bound on the step size in this case. 
To the best of our knowledge, this is the first plug-and-play convergence result for  superresolution and inpainting using standard NLM denoiser.  We conclude the paper with the following questions:\\
\indent (i) Similar to Theorem \ref{thm:inpainting_bound}, can an upper bound be computed for other linear inverse problems?\\
\indent (ii) Is it possible to extend the convergence analysis for FISTA  \cite{Beck2009_FISTA} and ADMM \cite{Boyd2011_ADMM}?

\section{Proofs}
\label{sec:proofs}

In this section, we give the detailed proofs of  Proposition \ref{prop:inpainting_Q} and Theorems \ref{thm:main} and  \ref{thm:inpainting_bound}.

\begin{proof}[\underline{Proof of Proposition \ref{prop:inpainting_Q}}]

Let $\Gamma$ be the set of observed pixels ($\lvert \Gamma \rvert < n$).
Then $\A^\top \! \A \in \mathbb{R}^{n \times n}$ is a diagonal matrix whose $i$-th diagonal element is $1$ if $i \in \Gamma$, and is $0$ if $i \notin \Gamma$. Note that $i$ denotes the linear index of a pixel w.r.t. to its vectorized representation, whereas $ \Gamma$ are the spatial locations.
Therefore the elements of $\Q = \W \A^\top \! \A \in \mathbb{R}^{n \times n}$ are given by $Q_{ij} = W_{ij}$ if $j \in \Gamma$, and $Q_{ij} = 0$ if $j \notin \Gamma$.
Thus, for $1 \leqslant i \leqslant n$, the left side of \eqref{eq:Q} becomes $0$ (since $W_{ij}=0$ for $j \notin \Omega_i$ by Assumption \ref{asm:W}), whereas the right side becomes
\begin{equation*}
\sum_{j \in \Omega_i \cap \Gamma} \!\!\!\! Q_{ij} \ \  + \! \!  \sum_{j \in \Omega_i \cap \Gamma^c} \!\!\!\!\! Q_{ij}  \ \ = \! \! \! \sum_{j \in \Omega_i \cap \Gamma} \!\!\!\! W_{ij}.
\end{equation*}
The sum on the right is positive since $W_{ij} > 0$ for $j \in \Omega_i$ (by Assumption \ref{asm:W}), and $\Omega_i \cap \Gamma$ is non-empty by the hypothesis of Proposition \ref{prop:inpainting_Q}.
Hence, \eqref{eq:Q} holds.
\end{proof}

The proof of Theorem \ref{thm:main} uses Gershgorin's disk theorem \cite[Chapter 7]{Meyer2000_MA}.
We state this result as applied to matrix $\P$. For $1 \leqslant i \leqslant n$, the Gershgorin disk $G_i$ is defined to be the disk in the complex plane with center $P_{ii}$ and radius $R_i = \sum_{j : j \neq i} |P_{ij}|$, i.e. $R_i$ is the sum of the unsigned  non-diagonal entries in the $i$th row. That is, $G_i = \left\{ z \in \mathbb{C} : | z - P_{ii} | \leqslant R_i \right\}$. Gershgorin's theorem states that every eigenvalue of $\P$ is contained in at least one of the Gershgorin disks $G_1,\ldots,G_n$.

\begin{proof}[\underline{Proof of Theorem \ref{thm:main}}]
We will show that there exists $\delta > 0$ such that all the Gershgorin disks $G_1,\ldots,G_n$ of $\P$ are contained in the interior of the unit disk in $\mathbb{C}$ (disk centered at 0 with radius $1$), provided that $\gamma \in (0,\delta)$.
This will imply that $\varrho(\P) < 1$ by Gershgorin's theorem, and hence convergence of $(\x_k)$ by Theorem \ref{thm:conv}.

Since $P_{ii}$ is real, the center of each disk $G_i$ is on the real axis. Hence, it is enough to show that $P_{ii} + R_i < 1$ and $P_{ii} - R_i > -1$, i.e. $P_{ii} \in (R_i-1,1-R_i)$; see Figure \ref{fig:Gershgorin}.
In this regard, we will prove the existence of $\alpha, \beta > 0$ such that, for all $1 \leqslant i \leqslant n$, the following hold:
(1) $P_{ii} > R_i - 1$  if $\gamma \in (0,\alpha)$; (2) $P_{ii} < 1 - R_i$ if $\gamma \in (0,\beta)$.
Recall that $\P$ is a function of $\gamma$. The theorem follows immediately if we set $\delta = \min( \alpha,\beta)$.

To prove the first result, we note that
\begin{equation*}
R_i - P_{ii} = \sum_{j : j \neq i} | W_{ij} - \gamma Q_{ij} | - (W_{ii} - \gamma Q_{ii}).
\end{equation*}
At $\gamma = 0$, this has the value $1 - 2 W_{ii}$, which is strictly less than $1$ since $W_{ii} >0$ (by Assumption \ref{asm:W} and the row-stochasticity of $\W$).
Moreover, $R_i - P_{ii}$ is a continuous function of $\gamma$ (note that $| \cdot |$ denotes absolute value of real numbers in the above equation). As a result, by continuity, there exists $\alpha_i >0$ such that   $R_i - P_{ii} < 1$ for $\gamma \in (0,\alpha_i )$.
The result now follows by letting $\alpha = \min(\alpha_1, \ldots, \alpha_n)$.

\begin{figure}
\centering
\includegraphics[width=0.46\linewidth]{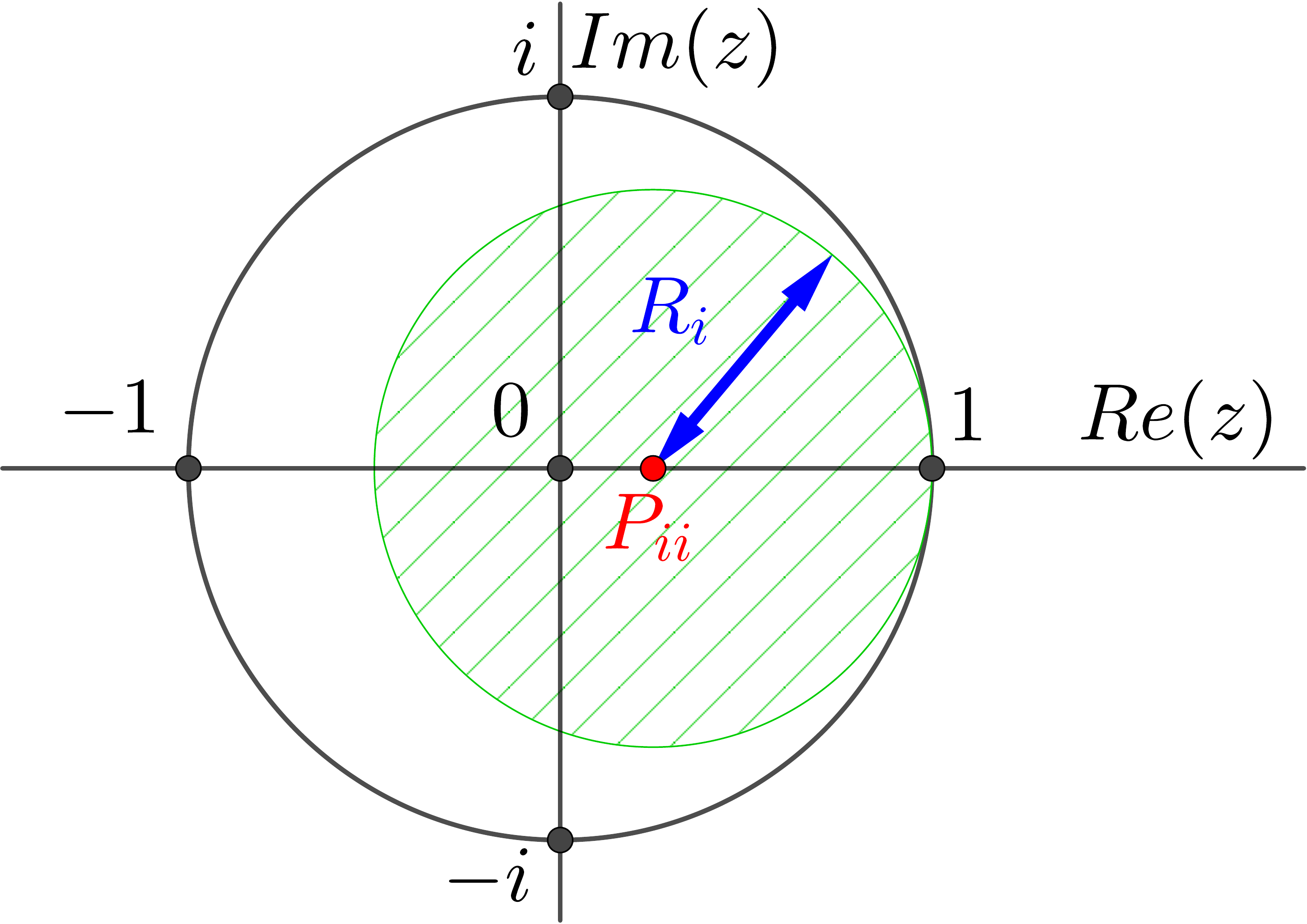}
\caption{Illustration for the proof of Theorem \ref{thm:main} which relies on Gershgorin's theorem \cite{Meyer2000_MA}. Every Gershgorin disk of $\P$ (shaded) is required to be in the interior of the unit circle in the complex plane. Since $P_{ii}$ is real, this is equivalent to the conditions $P_{ii} + R_i < 1$ and $P_{ii} - R_i > -1$.}
\label{fig:Gershgorin}
\end{figure}

To prove the second result, we note that 
\begin{align}
P_{ii} &+ R_i = W_{ii} - \gamma Q_{ii} + \sum_{j : j \neq i} | W_{ij} -\gamma Q_{ij} | \nonumber \\
&= W_{ii} - \gamma Q_{ii} + \! \! \! \! \sum_{j \neq i, W_{ij} = 0} \! \! \! \! \! \! \gamma Q_{ij} + \! \! \! \! \sum_{j \neq i, W_{ij} > 0} \! \! \! \! \! \! | W_{ij} - \gamma Q_{ij} |, \label{eq:thm1}
\end{align}
where we used the fact that $Q_{ij} \geqslant 0$ (by Assumptions \ref{asm:W} and \ref{asm:A}) in the first summation to drop the absolute value symbols. 
Moreover, we will show that we can also drop the absolute value symbols in the last summation in \eqref{eq:thm1}.
This is because the last summation is over the set $\{ j \neq i :W_{ij} > 0 \}$.
Since the map $\gamma \mapsto W_{ij} - \gamma Q_{ij}$ is continuous, for every $j$ in this set there exists $\beta_{ij} > 0$ such that $W_{ij} - \gamma Q_{ij} > 0$ if $0 \leqslant \gamma < \beta_{ij}$.
Let $\beta_i = \min \{\beta_{ij} : j \neq i, W_{ij} > 0 \}$.
Since $\beta_i$ is the minimum of a finite set of positive numbers, it is positive.
As a result, by setting $\gamma < \beta_i$, we can drop the absolute values in the last summation in \eqref{eq:thm1}.
Thus, for $0 \leqslant \gamma < \beta_i$, we obtain that
\begin{align}
P_{ii} + R_i &= \Big( W_{ii} + \! \! \! \! \sum_{\substack{j \in  \Omega_i , \\ j \neq i}} W_{ij} \Big)
+  \gamma \! \! \! \! \sum_{\substack{j \notin \Omega_i, \\ j \neq i}} Q_{ij}
- \gamma  \Big( Q_{ii} +  \! \! \! \! \sum_{\substack{j \in \Omega_i, \\ j \neq i}} Q_{ij} \Big) \nonumber \\
&= 1 + \gamma \Big( \sum_{j \notin \Omega_i} Q_{ij}  - \sum_{j \in  \Omega_i} Q_{ij} \Big), \label{eq:thm2}
\end{align}
where we have used Assumption \ref{asm:W} and the fact that that the row-sum $\sum_{j} W_{ij}$ is $1$, to simplify the terms.
From \eqref{eq:thm2}, we see that $P_{ii} + R_i$ is a linear function of $\gamma$ for $0 \leqslant \gamma < \beta_i$, with slope 
\begin{equation*}
\sum_{j \notin \Omega_i} Q_{ij} - \sum_{j \in \Omega_i} Q_{ij}.
\end{equation*}
By Assumption \ref{asm:Q}, this slope is negative. Hence, $P_{ii} + R_i$ is strictly decreasing in $\gamma$ for $0 \leqslant \gamma < \beta_i$.
Moreover, the value of $P_{ii} + R_i $ at $\gamma = 0$ is $1$.
Hence, $P_{ii} + R_i < 1$ for $0 < \gamma < \beta_i$.
The result now follows by letting $\beta = \min ( \beta_1,\ldots,\beta_n )$.
\end{proof}

\begin{proof}[\underline{Proof of Theorem \ref{thm:inpainting_bound}}]
This is similar to the proof of Theorem \ref{thm:main}.
However, as stated in the proof of Proposition \ref{prop:inpainting_Q}, $Q_{ij}$ is either $W_{ij}$ or $0$  for inpainting.
Using this, it is not difficult to verify that we can take $\alpha_i = 1$ and $\beta_{ij} = 1$  in the proof of Theorem \ref{thm:main}.
As a result, we get $\alpha = \beta = 1$, so that  $\delta =\min(\alpha,\beta) = 1$.
\end{proof}

\bibliographystyle{IEEEtran}
\bibliography{citations}

\newpage
\onecolumn
\noindent\textbf{\Large{Supplementary material}}

\begin{figure}[H]
\centering
\subfloat[Ground-truth.]{\includegraphics[width=0.15\linewidth]{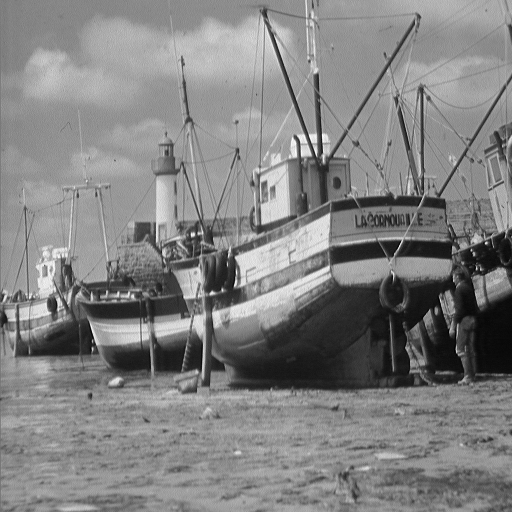}}
\hspace{0.1mm}
\subfloat[Observed pixels.]{\includegraphics[width=0.15\linewidth]{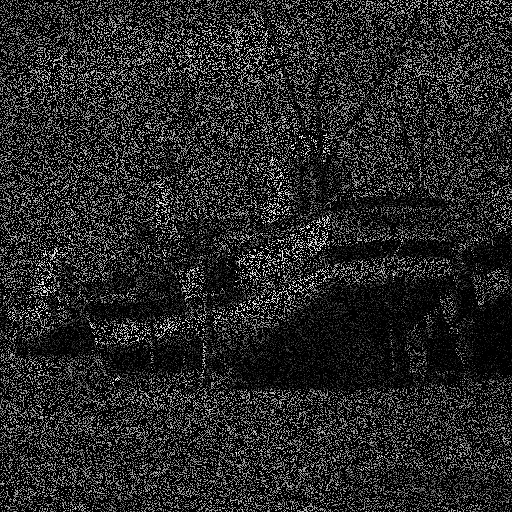}}
\hspace{0.1mm}
\subfloat[Output ($25.5$ dB).]{\includegraphics[width=0.15\linewidth]{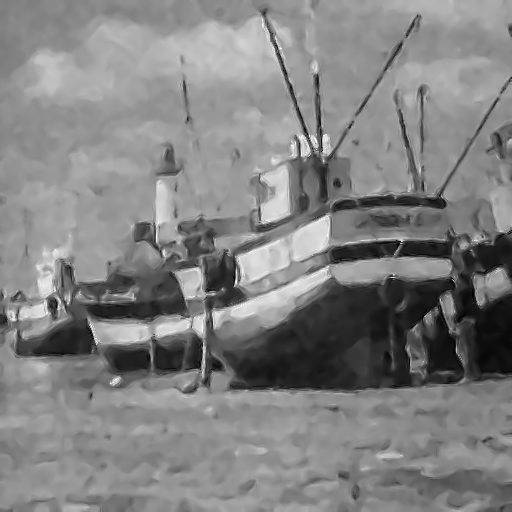}}
\hspace{0.1mm}
\subfloat[PSNR (dB).]{\includegraphics[width=0.225\linewidth]{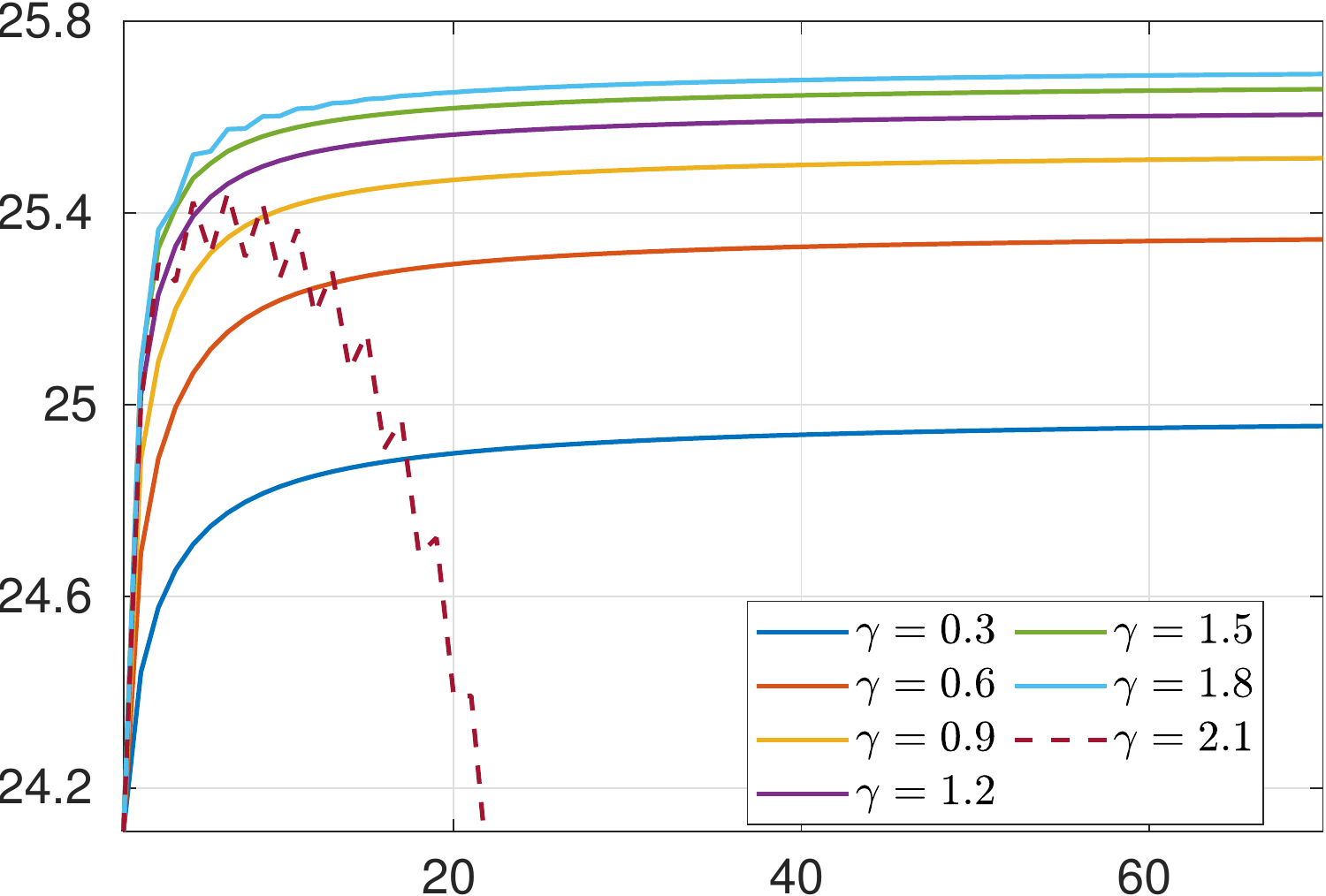}}
\hspace{0.1mm}
\subfloat[Residual (log scale).]{\includegraphics[width=0.219\linewidth]{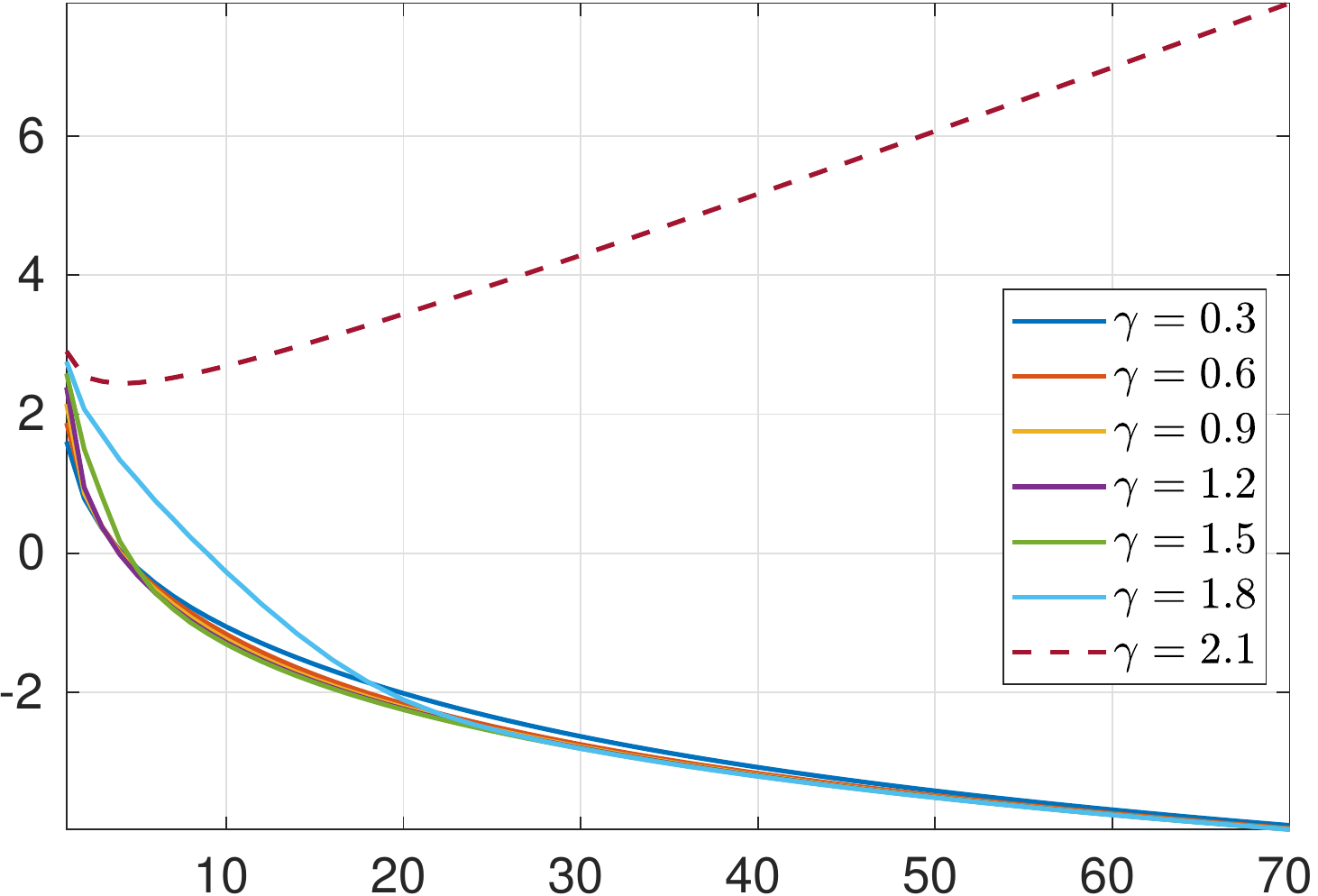}}
\caption{Image inpainting with $70 \%$ missing pixels and Gaussian noise (standard deviation $20/255$). We have used PnP-ISTA with NLM denoiser. The result in (c) is using $\gamma = 0.9$.}
\label{fig:inpainting_supp}
\end{figure}

\begin{figure}[H]
\centering
\subfloat[Ground-truth.]{\includegraphics[width=0.15\linewidth]{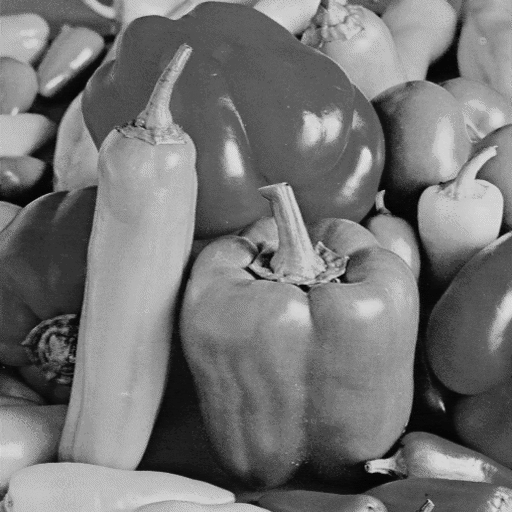}}
\hspace{0.1mm}
\subfloat[Blurred.]{\includegraphics[width=0.15\linewidth]{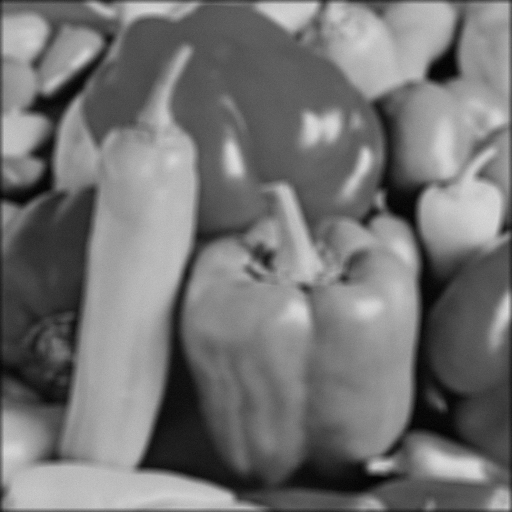}}
\hspace{0.1mm}
\subfloat[Output ($30.2$ dB).]{\includegraphics[width=0.15\linewidth]{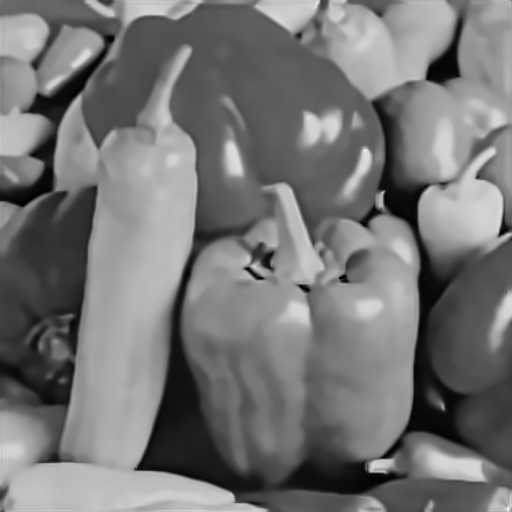}}
\hspace{0.1mm}
\subfloat[PSNR (dB).]{\includegraphics[width=0.22\linewidth]{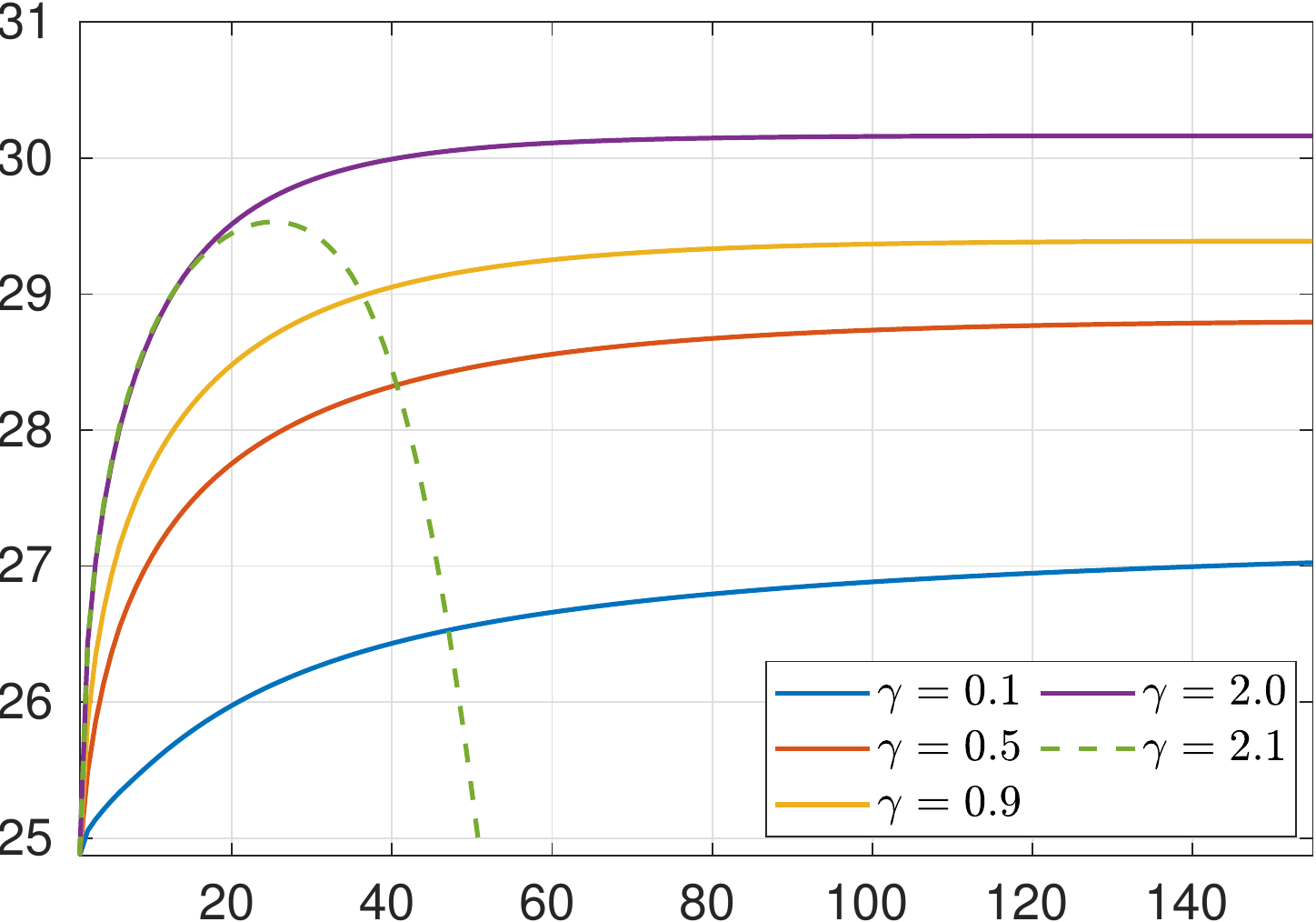}}
\hspace{0.1mm}
\subfloat[Residual (log scale).]{\includegraphics[width=0.22\linewidth]{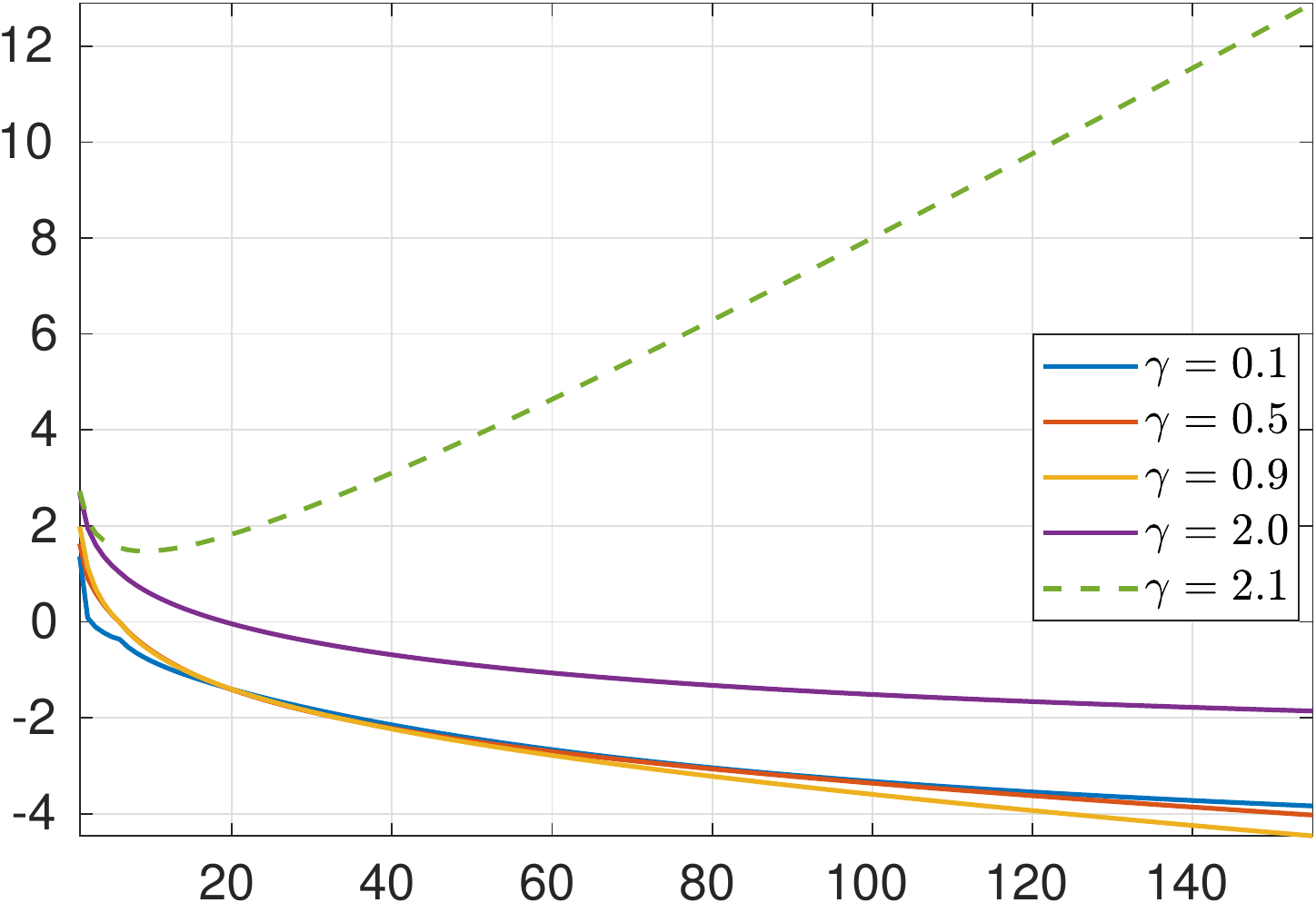}}
\caption{Image deblurring with a uniform blurring kernel and Gaussian noise with standard deviation $2/255$. We have used PnP-ISTA with NLM denoiser. The  image in (c) is for $\gamma = 2$.}
\label{fig:deblurring_supp}
\end{figure}

\begin{figure}[H]
\centering
\subfloat[Output ($25.5$ dB).]{\includegraphics[width=0.15\linewidth]{./supp_boat_out.png}}
\hspace{0.1mm}
\subfloat[Output ($23.9$ dB).]{\includegraphics[width=0.15\linewidth]{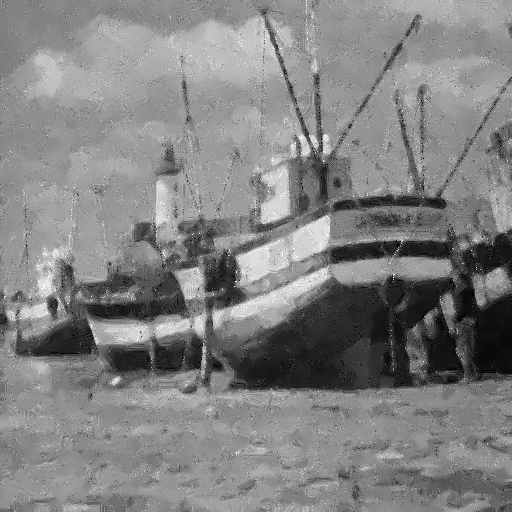}}
\hspace{0.1mm}
\subfloat[PSNR (dB).]{\includegraphics[width=0.225\linewidth]{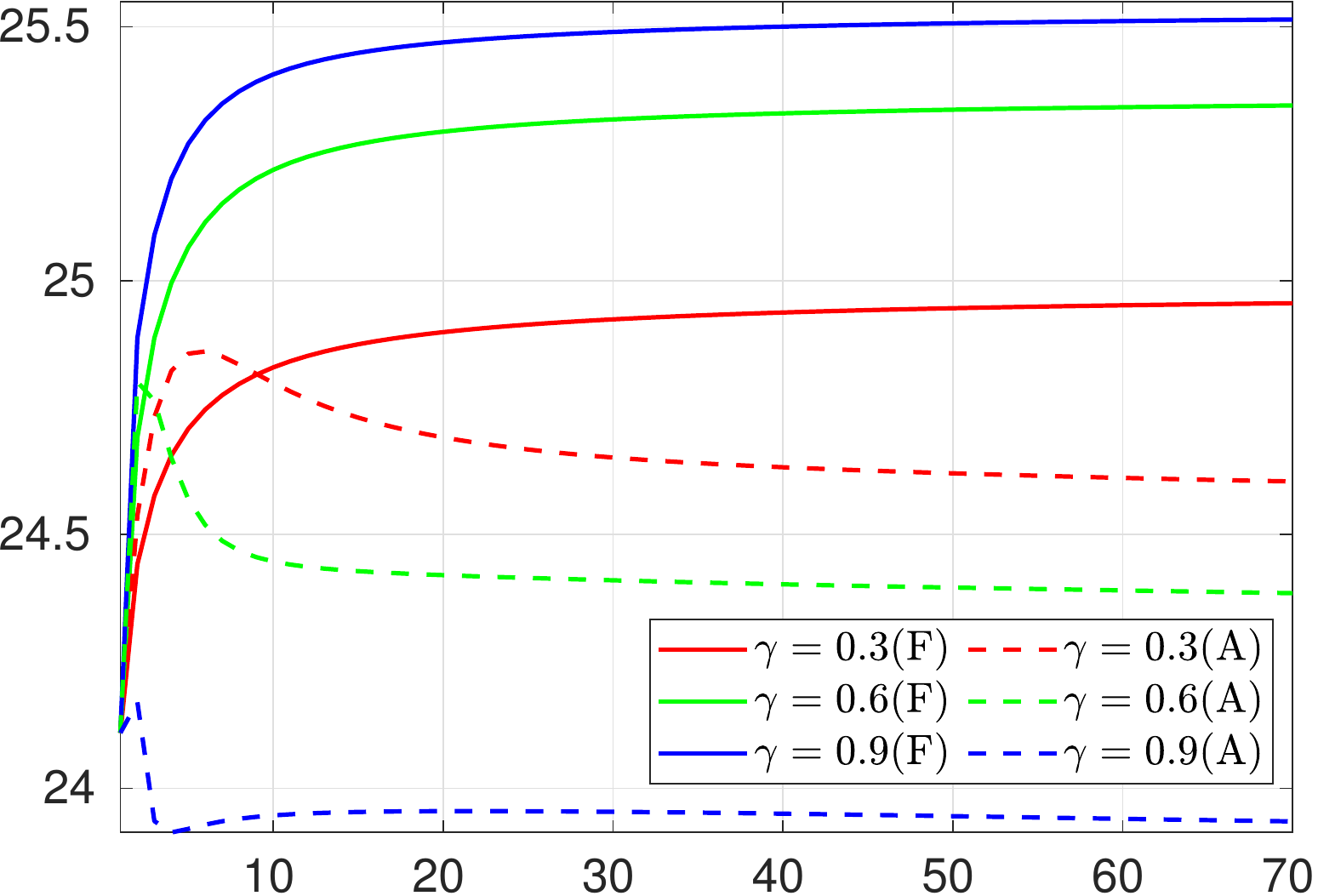}}
\hspace{0.1mm}
\subfloat[Residual (log scale).]{\includegraphics[width=0.216\linewidth]{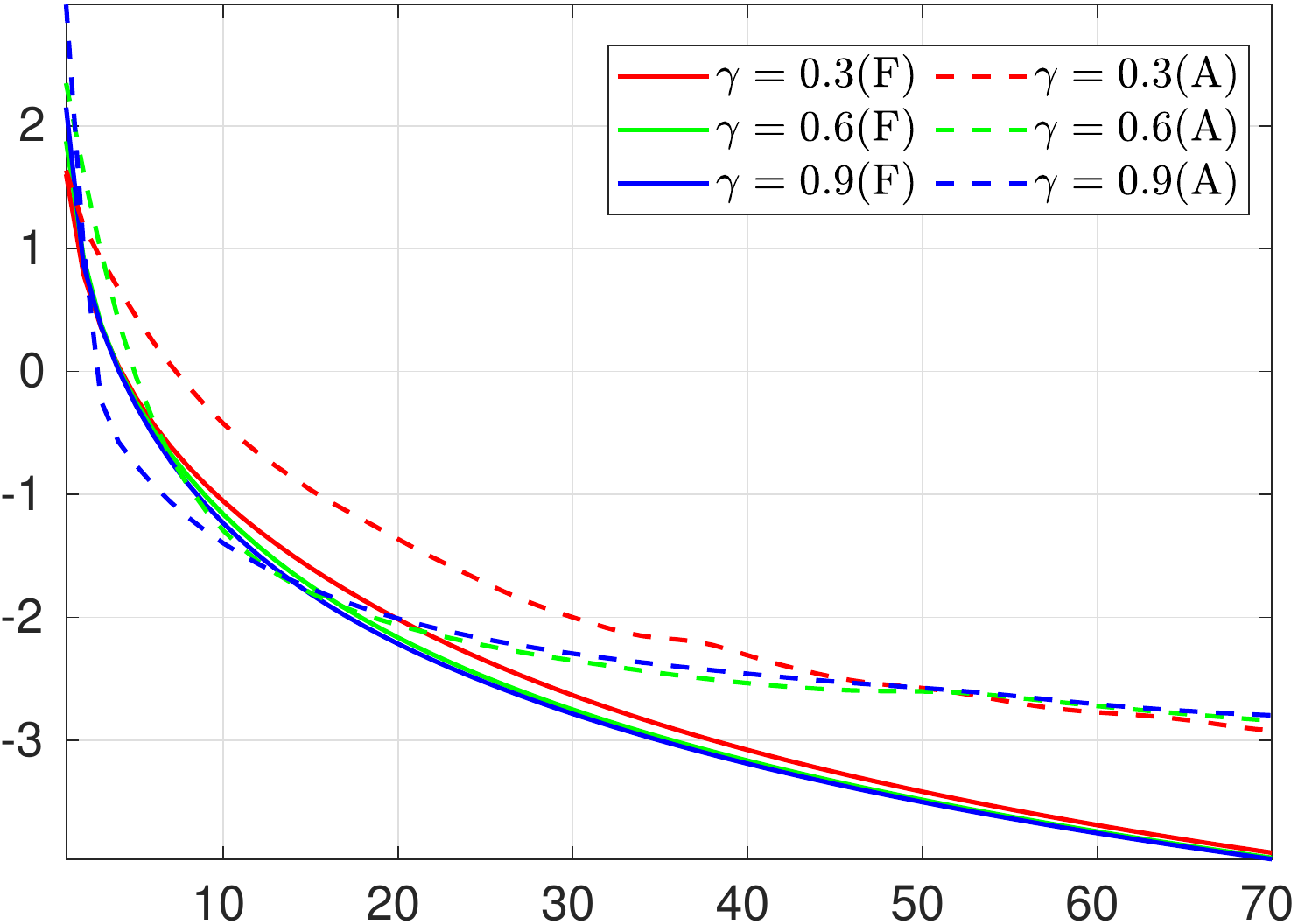}}
\caption{ Image inpainting using (a) fixed and (b) iteration-adaptive NLM denoiser with $\gamma = 0.9$. The plots show the evolution of PSNR and residual in the two cases (F: Fixed, A: Adaptive) for different step sizes. The input image is the same as in Figure \ref{fig:inpainting}.}
\label{fig:adaptive_supp}
\end{figure}

Here, we report a couple of additional results for inpainting  and deblurring using PnP-ISTA with NLM denoiser.
The inpainting experiment (Figure \ref{fig:inpainting_supp}) was performed at high noise level (standard deviation $20/255$).
As in the experiment reported in the main paper, we observe that the algorithm converges not only for $\gamma < 1$ (which is expected, by Theorem 5 in the paper), but also for higher values up to $\gamma = 2$.
However, it seems to diverge for $\gamma > 2$.
A similar observation holds for the deblurring experiment in Figure \ref{fig:deblurring_supp}.
Although convergence in the case of deblurring cannot be  guaranteed (since it is not known whether Assumption 2(iii) holds for deblurring), the iterates seem to converge for $\gamma \leq 2$ and diverge for $\gamma > 2$.

Finally, for completeness, we perform an inpainting experiment (Figure \ref{fig:adaptive_supp}) in which we compare the performance of PnP-ISTA across two settings.
In the first setting, the NLM denoiser $\W$ is linear (this was our assumption throughout the paper), where the weights are computed from a fixed image.
In the second setting, the NLM weights in a given iteration are computed from the output image from the previous iteration. The denoiser  is no longer linear in this case.
It can be observed from Figure \ref{fig:adaptive_supp} that using a fixed (linear) denoiser tends to result in better outputs.
We have used the same parameter settings for NLM in both cases (and in Figure \ref{fig:inpainting_supp}), and the parameters are not changed across iterations.
Interestingly,  PnP-ISTA seems to stabilize even with changing weights, although the analysis in the paper is not applicable in this case.

\end{document}